\documentclass[12pt]{article}
\usepackage{graphicx}
\usepackage{epsfig}
\usepackage{epstopdf}
\DeclareGraphicsExtensions{.pdf,.eps,.png,.jpg,.mps}
\usepackage{caption}
\usepackage{float}

\usepackage{url}
\usepackage[implicit=false]{hyperref}
\usepackage{cite}

\def\lsim{\mathrel{\rlap{\lower3pt\hbox{\hskip0pt$\sim$}}
     \raise1pt\hbox{$<$}}}         
\def\gsim{\mathrel{\rlap{\lower4pt\hbox{\hskip1pt$\sim$}}
     \raise1pt\hbox{$>$}}}         

\usepackage{amsmath}
\usepackage{amsfonts}

\begin{document}
\begin{titlepage}

\centerline{\Large \bf Machine Learning Risk Models}
\medskip

\centerline{Zura Kakushadze$^\S$$^\dag$\footnote{\, Zura Kakushadze, Ph.D., is the President of Quantigic$^\circledR$ Solutions LLC,
and a Full Professor at Free University of Tbilisi. Email: \href{mailto:zura@quantigic.com}{zura@quantigic.com}} and Willie Yu$^\sharp$\footnote{\, Willie Yu, Ph.D., is a Research Fellow at Duke-NUS Medical School. Email: \href{mailto:willie.yu@duke-nus.edu.sg}{willie.yu@duke-nus.edu.sg}}}
\bigskip

\centerline{\em $^\S$ Quantigic$^\circledR$ Solutions LLC}
\centerline{\em 1127 High Ridge Road \#135, Stamford, CT 06905\,\,\footnote{\, DISCLAIMER: This address is used by the corresponding author for no
purpose other than to indicate his professional affiliation as is customary in
publications. In particular, the contents of this paper
are not intended as an investment, legal, tax or any other such advice,
and in no way represent views of Quantigic$^\circledR$ Solutions LLC,
the website \url{www.quantigic.com} or any of their other affiliates.
}}
\centerline{\em $^\dag$ Free University of Tbilisi, Business School \& School of Physics}
\centerline{\em 240, David Agmashenebeli Alley, Tbilisi, 0159, Georgia}
\centerline{\em $^\sharp$ Centre for Computational Biology, Duke-NUS Medical School}
\centerline{\em 8 College Road, Singapore 169857}
\medskip
\centerline{(January 1, 2019)}

\bigskip
\medskip

\begin{abstract}
{}We give an explicit algorithm and source code for constructing risk models based on machine learning techniques. The resultant covariance matrices are not factor models. Based on empirical backtests, we compare the performance of these machine learning risk models to other constructions, including statistical risk models, risk models based on fundamental industry classifications, and also those utilizing multilevel clustering based industry classifications.
\end{abstract}
\medskip
\end{titlepage}

\newpage

\section{Introduction and Summary}

{}In most practical quant trading applications\footnote{\, Similar issues are also present in other practical applications unrelated to trading or finance.} one faces an old problem when computing a sample covariance matrix of returns: the number $N$ of returns (e.g., the number of stocks in the trading universe) is much larger than the number $T$ of observations in the time series of returns. The sample covariance matrix $C_{ij}$ ($i,j=1,\dots,N$) in this case is badly singular: its rank is at best $T-1$. So, it cannot be inverted, which is required in, e.g., mean-variance optimization \cite{Markowitz1952}. In fact, the singularity of $C_{ij}$ is only a small part of the trouble: its off-diagonal elements (more precisely, sample correlations) are notoriously unstable out-of-sample.

{}The aforesaid ``ills" of the sample covariance matrix are usually cured via multifactor risk models,\footnote{\, For a general discussion, see, e.g., \cite{GrinoldKahn}. For explicit implementations (including source code), see, e.g., \cite{HetPlus}, \cite{StatRM}.} where stock returns are (linearly) decomposed into contributions stemming from some number $K$ of common underlying factors plus idiosyncratic ``noise" pertaining to each stock individually. This is a way of dimensionally reducing the problem in that one only needs to compute a factor covariance matrix $\Phi_{AB}$ ($A,B=1,\dots,K$), which is substantially smaller than $C_{ij}$ assuming $K\ll N$.\footnote{\, This does not solve all problems, however. Thus, unless $K < T$, the sample factor covariance matrix is still singular (albeit the model covariance matrix $\Gamma_{ij}$ that replaces $C_{ij}$ need not be). Furthermore, the out-of-sample instability is still present in sample factor correlations. This can be circumvented via the heterotic risk model construction \cite{Het}; see below.}

{}In statistical risk models\footnote{\, See \cite{StatRM}, which gives complete source code, and references therein.} the factors are based on the first $K$ principal components of the sample covariance matrix $C_{ij}$ (or the sample correlation matrix).\footnote{\, The (often misconstrued) ``shrinkage" method \cite{Ledoit} is nothing but a special type of statistical risk models; see \cite{S=FM}, \cite{StatRM} for details.} In this case the number of factors is limited ($K \leq T-1$), and, furthermore, the principal components beyond the first one are inherently unstable out-of-sample. In contrast, factors based on a granular fundamental industry classification\footnote{\, E.g., BICS (Bloomberg Industry Classification System), GICS (Global Industry Classification Standard), ICB (Industry Classification Benchmark), SIC (Standard Industrial Classification), etc.} are much more ubiquitous (in hundreds), and also stable, as stocks seldom jump industries. Heterotic risk models \cite{Het} based on such industry classifications sizably outperform statistical risk models.\footnote{\, In the heterotic risk model construction the sample factor covariance matrix at the most granular level in the industry classification typically would be singular. However, this is rectified by modeling the factor covariance matrix by another factor model with factors based on the next-less-granular level in the industry classification, and this process of dimensional reduction is repeated until the resultant factor covariance matrix is small enough to be nonsingular and sufficiently stable out-of-sample \cite{Het}, \cite{HetPlus}. Here one can also include non-industry style factors. However, their number is limited (especially for short horizons) and, contrary to an apparent common misconception, style factors generally are poor proxies for modeling correlations and add little to no value \cite{HetPlus}.} Another alternative is to replace the fundamental industry classification in the heterotic risk model construction by a statistical industry classification based on clustering (using machine learning techniques) the return time series data \cite{StatIC},\footnote{\, Such statistical industry classifications can be multilevel and granular.} without any reference to a fundamental industry classification. Risk models based on statistical industry classifications outperform statistical risk models but underperform risk models based on fundamental industry classifications \cite{StatIC}.

{}In this paper we discuss a different approach to building a risk model using machine learning techniques. The idea is simple. A sample covariance matrix $C_{ij}$ is singular (assuming $T \ll N$), but it is semi-positive definite. Imagine that we could compute a large number $M$ of ``samplings" of $C_{ij}$, call them $C_{ij}^{(m)}$, $m=1,\dots,M$, where each ``sampling" is semi-positive definite. Consider their mean\footnote{\, In fact, instead of the arithmetic mean, here we can more generally consider a weighted average with some positive weights $w_m$ (see below). Also, in this paper $C_{ij}^{(m)}$ are nonsingular.}
\begin{equation}\label{samplings}
 \Gamma_{ij} = {1\over M} \sum_{m=1}^M C_{ij}^{(m)}
\end{equation}
By construction $\Gamma_{ij}$ is semi-positive definite. In fact, assuming $C_{ij}^{(m)}$ are all (sizably) different from each other, $\Gamma_{ij}$ generically will be positive definite and invertible (for large enough $M$). So, the idea is sound, at least superfluously, but the question is, what should these ``samplings" $C_{ij}^{(m)}$ be? Note that each element of the sample covariance matrix $C_{ij}$ ($i\neq j$) only depends on the time series of the corresponding two stock returns $R_i(t)$ and $R_j(t)$, and not on the universe of stocks, so any cross-sectional ``samplings" cannot be based on sample covariance matrices. In principle, serial ``samplings" could be considered if a long history were available. However, here we assume that our lookback is limited, be it due to a short history that is available, or, more prosaically, due to the fact that data from a while back is not pertinent to forecasting risk for short horizons as market conditions change.

{}A simple way around this is to consider cross-sectional ``samplings" $C_{ij}^{(m)}$ that are not sample covariance matrices but are already dimensionally reduced, even though they do not have to be invertible. Thus, given a clustering of $N$ stocks into $K$ clusters, we can build a multifactor risk model, e.g., via an incomplete heterotic construction (see below). Different clusterings then produce different ``samplings" $C_{ij}^{(m)}$, which we average via Eq. (\ref{samplings}) to obtain a positive definite $\Gamma_{ij}$, which is {\em not} a factor model. However, as usual, the devil is in the detail, which we discuss in Section \ref{sec2}. E.g., the matrix (\ref{samplings}) can have nearly degenerate or small eigenvalues, which requires further tweaking $\Gamma_{ij}$ to avert, e.g., undesirable effects on optimization.

{}In Section \ref{sec3} we discuss backtests to compare the machine learning risk models of this paper to statistical risk models, and heterotic risk models based on fundamental industry classification and statistical industry classification. We briefly conclude in Section \ref{sec4}. Appendix \ref{app.A} provides R source code\footnote{\, The code in Appendix A is not written to be ``fancy" or optimized for speed or otherwise.} for machine learning risk models, and some important legalese relating to this code is relegated to Appendix \ref{app.B}.

\section{Heterotic Construction and Sampling}\label{sec2}

{}So, we have time series of returns (say, daily close-to-close returns) $R_{is}$ for our $N$ stocks ($i=1,\dots,N$, $s=1,\dots, T$, and $s=1$ corresponds to the most recent time in the time series). Let us assume that we have a clustering of our $N$ stocks into $K$ clusters, where $K$ is sizably smaller than $N$, and each stock belongs to one and only one cluster. Let the clusters be labeled by $A=1,\dots,K$. So, we have a map
\begin{eqnarray}\label{G.map}
 &&G:\{1,\dots,N\}\mapsto\{1,\dots,K\}
\end{eqnarray}
Following \cite{Het}, we can model the sample correlation matrix $\Psi_{ij} = C_{ij}/\sigma_i\sigma_j$ (here $\sigma_i^2 = C_{ii}$ are the sample variances) via a factor model:
\begin{eqnarray}\label{model.cor}
 &&{\widetilde \Psi}_{ij} = \xi_i^2~\delta_{ij} + \sum_{A,B = 1}^K \Omega_{iA}~\Phi_{AB}~\Omega_{jB} = \xi_i^2~\delta_{ij} + U_i~U_j~\Phi_{G(i), G(j)}\\
 &&\Omega_{iA} = U_i~\delta_{G(i), A}\\
 &&\xi_i^2 = 1 - \lambda(G(i))~U_i^2\\
 &&\Phi_{AB} = \sum_{i\in J(A)} \sum_{j\in J(B)} U_i~\Psi_{ij}~U_j\label{Phi}
\end{eqnarray}
Here the $N_A$ components of $U_i$ for $i\in J(A)$ are given by the first principal component of the $N(A)\times N(A)$ matrix $[\Psi(A)]_{ij} = \Psi_{ij}$, $i,j\in J(A)$, where $J(A) =\{i|G(i) = A\}$ is the set of the values of the index $i$ corresponding to the cluster labeled by $A$, and $N_A = |J(A)|$ is the number of such $i$. Also, $\lambda(A)$ is the largest eigenvalue (corresponding to the first principal component) of the matrix $[\Psi(A)]_{ij}$. The matrix $\Omega_{iA}$ is the factor loadings matrix, $\xi_i^2$ is the specific variance, and the factor covariance matrix $\Phi_{AB}$ has the property that $\Phi_{AA} = \lambda(A)$. By construction, ${\widetilde \Psi}_{ii} = 1$, and the matrix ${\widetilde \Psi}_{ij}$ is positive-definite. However, $\Phi_{AB}$ is singular unless $K \leq T - 1$.

{}This is because the rank of $\Psi_{ij}$ is (at most) $T-1$. Let $V_i^{(a)}$ be the principal components of $\Psi_{ij}$ with the corresponding eigenvalues $\lambda^{(a)}$ ordered decreasingly ($a=1,\dots,N$). More precisely, at most $T-1$ eigenvalues $\lambda^{(a)}$, $a=1,\dots,T-1$ are nonzero, and the others vanish. So, we have
\begin{eqnarray}\label{fac.cov}
 &&\Phi_{AB} = \sum_{a=1}^{T-1}\lambda^{(a)}~{\widetilde U}_A^{(a)}~{\widetilde U}_B^{(a)}\\
 &&{\widetilde U}_A^{(a)} = \sum_{i\in J(A)} U_i~V_i^{(a)}
\end{eqnarray}
So, the rank of $\Phi_{AB}$ is (at most) $T-1$, and the above incomplete heterotic construction provides a particular regularization of the statistical risk model construction based on principal components. In the complete heterotic construction $\Phi_{AB}$ itself is modeled via another factor model, and this nested ``Russian-doll" embedding is continued until at the final step the factor covariance matrix (which gets smaller and smaller at each step) is nonsingular (and sufficiently stable out-of-sample).

\subsection{Sampling via Clustering}

{}However, there is another way, which is what we refer to as ``machine learning risk models" here. Suppose we have $M$ different clusterings. Let ${\widetilde \Psi}_{ij}^{(m)}$ be the model correlation matrix (\ref{model.cor}) for the $m$-th clustering ($m = 1,\dots, M$). Then we can construct a model correlation matrix as a weighted sum
\begin{eqnarray}
 &&{\widetilde\Psi}_{ij} = \sum_{m=1}^M w_m~{\widetilde \Psi}_{ij}^{(m)}\\
 &&\sum_{m=1}^M w_m = 1
\end{eqnarray}
The simplest choice for the weights is to have equal weighting: $w_m = 1/M$. More generally, so long as the weights $w_m$ are positive, the model correlation matrix ${\widetilde\Psi}_{ij}$ is positive-definite. (Also, by construction ${\widetilde\Psi}_{ii} = 1$.) However, combining a large number $M$ of ``samplings" ${\widetilde \Psi}_{ij}^{(m)}$ accomplishes something else: each ``sampling" provides a particular regularization of the sample correlation matrix, and combining such samplings covers many more directions in the risk space than each individual ``sampling". This is because ${\widetilde U}_A^{(a)}$ in Eq. (\ref{fac.cov}) are different for different clusterings.

\subsection{K-means}

{}We can use k-means \cite{Forgy}, \cite{Lloyd1957}, \cite{Lloyd1982}, \cite{Hartigan}, \cite{HartWong}, \cite{MacQueen}, \cite{Steinhaus} for our clusterings. Since k-means is nondeterministic, it automatically produces a different ``sampling" with each run. The idea behind k-means is to partition $N$ observations into $K$ clusters such that each observation belongs to the cluster with the nearest mean. Each of the $N$ observations is actually a $d$-vector, so we have an $N \times d$ matrix $X_{is}$, $i=1,\dots,N$, $s=1,\dots,d$. Let $C_a$ be the $K$ clusters, $C_a = \{i| i\in C_a\}$, $a=1,\dots,K$. Then k-means attempts to minimize
\begin{equation}\label{k-means}
 g = \sum_{a=1}^K \sum_{i \in C_a} \sum_{s=1}^d  \left(X_{is} - Y_{as}\right)^2
\end{equation}
where
\begin{equation}\label{centers}
 Y_{as} = {1\over n_a} \sum_{i\in C_a} X_{is}
\end{equation}
are the cluster centers (i.e., cross-sectional means),\footnote{\, Throughout this paper ``cross-sectional" refers to ``over the index $i$".} and $n_a = |C_a|$ is the number of elements in the cluster $C_a$. In Eq. (\ref{k-means}) the measure of ``closeness" is chosen to be the Euclidean distance between points in ${\bf R}^d$, albeit other measures are possible.\footnote{\, E.g., the Manhattan distance, cosine similarity, etc.}

\subsection{What to Cluster?}

{}Here we are not going to reinvent the wheel. We will simply use the prescription of \cite{StatIC}. Basically, we can cluster the returns, i.e., take $X_{is} = R_{is}$ (then $d = T$). However, stock volatility is highly variable, and its cross-sectional distribution is not even quasi-normal but highly skewed, with a long tail at the higher end -- it is roughly log-normal. Clustering returns does not take this skewness into account and inadvertently we might be clustering together returns that are not at all highly correlated solely due to the skewed volatility factor. A simple ``machine learning" solution is to cluster the normalized returns ${\widetilde R}_{is} = R_{is} / \sigma_i$, where $\sigma_i^2 = \mbox{Var}(R_{is})$ is the serial variance ($\sigma_i^2 = C_{ii}$). However, as was discussed in detail in \cite{StatIC}, this choice would also be suboptimal and this is where quant trading experience and intuition trumps generic machine learning ``lore". It is more optimal to cluster ${\widehat R}_{is} = R_{is} / \sigma_i^2$ (see \cite{StatIC} for a detailed explanation). A potential practical hiccup with this is that if some stocks have very low volatilities, we could have large ${\widehat R}_{is}$ for such stocks. To avoid any potential issues with computations, we can ``smooth" this out via ``Winsorization" of sorts (MAD = mean absolute deviation):\footnote{\, This is one possible tweak. Others produce similar results.}
\begin{eqnarray}\label{tweak}
 &&{\widehat R}_{is} = {R_{is} \over {\sigma_i u_i}}\\
 &&u_i = {\sigma_i\over v}\\
 &&v = \exp(\mbox{Median}(\ln(\sigma_i)) - 3~\mbox{MAD}(\ln(\sigma_i)))
\end{eqnarray}
and for all $u_i < 1$ we set $u_i = 1$. This is the definition of ${\widehat R}_{is}$ that is used in the source code internally. Furthermore, Median($\cdot$) and MAD($\cdot$) above are cross-sectional.

\subsection{A Tweak}\label{sub.tail}

{}The number of clusters $K$ is a hyperparameter. In principle, it can be fixed by adapting the methods discussed in \cite{StatIC}. However, in the context of this paper, we will simply keep it as a hyperparameter and test what we get for its various values. As $K$ increases, in some cases it is possible to get relatively small eigenvalues in the model correlation matrix ${\widetilde\Psi}_{ij}$, or nearly degenerate eigenvalues. This can cause convergence issues in optimization with bounds (see below). To circumvent this, we can slightly deform ${\widetilde\Psi}_{ij}$ for such values of $K$.

\begin{figure}[ht]
\centering
\includegraphics[scale=1.0]{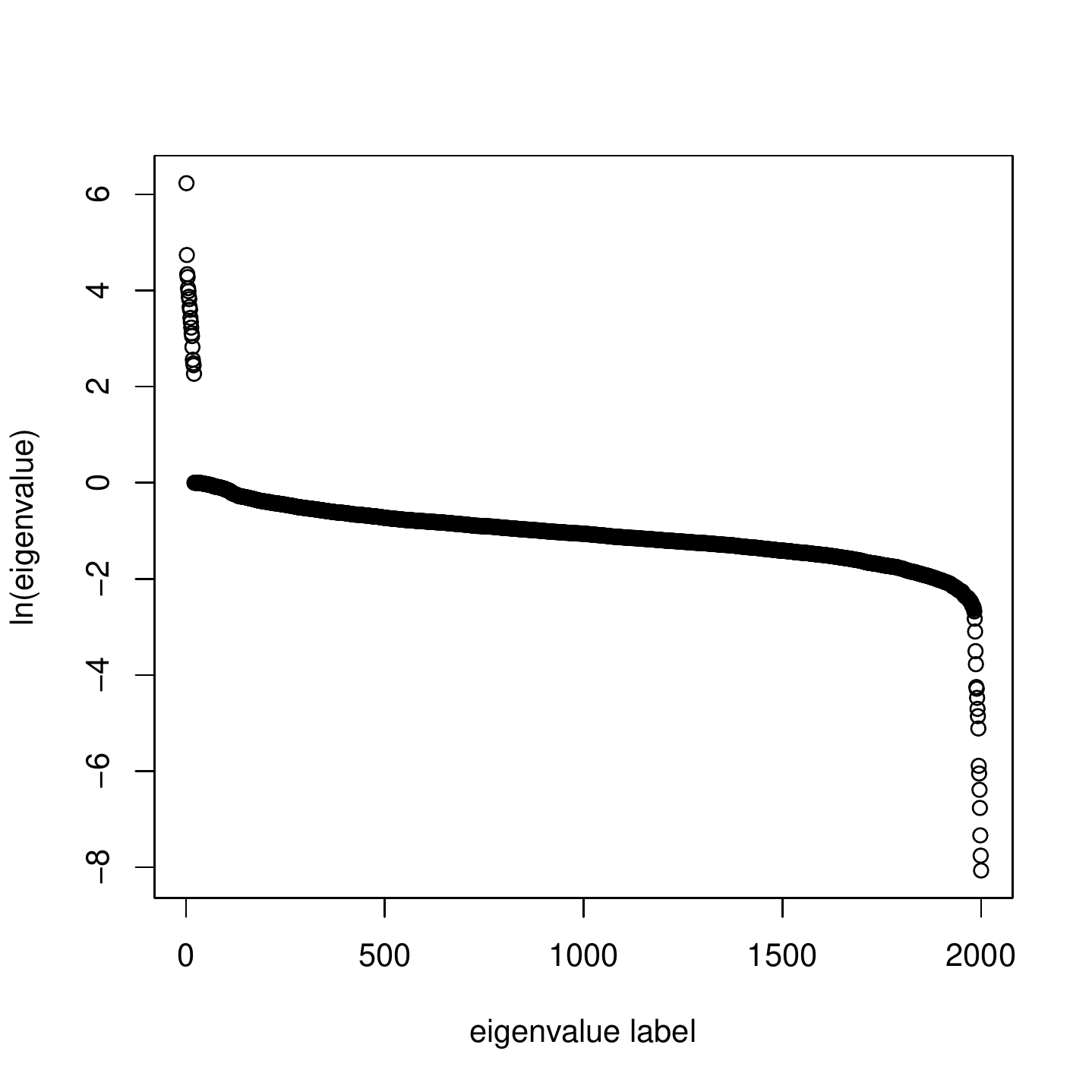}
\caption{A typical graph of the log of the eigenvalues (ordered decreasingly) of the model correlation matrix ${\widetilde\Psi}^{(m)}_{ij}$ for a single ``sampling" ($M=1$). The number of clusters $K=150$. See Subsection \ref{sub.tail} for details.}
\label{Figure1}
\end{figure}

\begin{figure}[ht]
\centering
\includegraphics[scale=1.0]{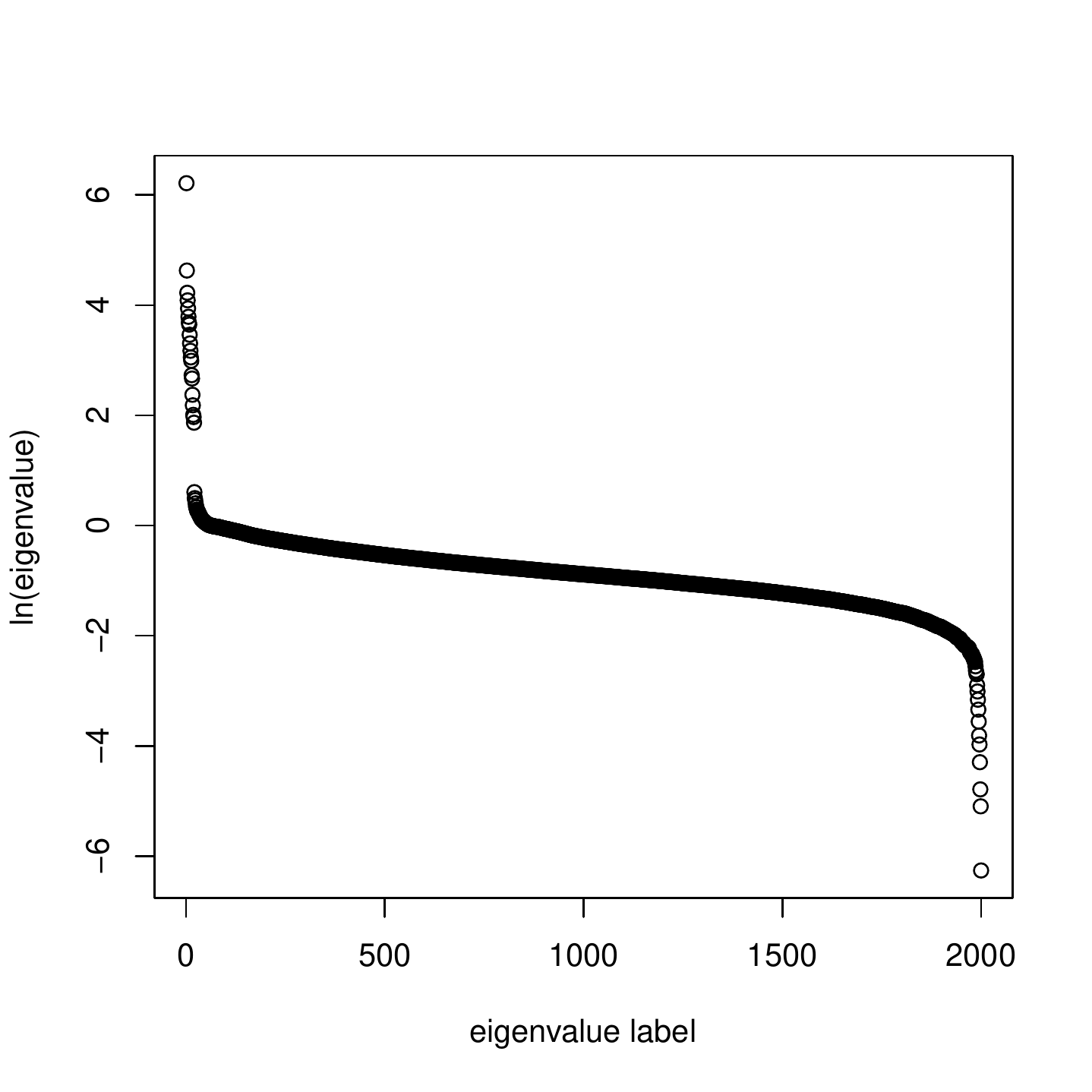}
\caption{A typical graph of the log of the eigenvalues (ordered decreasingly) of the model correlation matrix ${\widetilde\Psi}_{ij}$ obtained by combining $M=100$ ``samplings" (with equal weights). The number of clusters $K=150$. See Subsection \ref{sub.tail} for details.}
\label{Figure2}
\end{figure}

\begin{figure}[ht]
\centering
\includegraphics[scale=1.0]{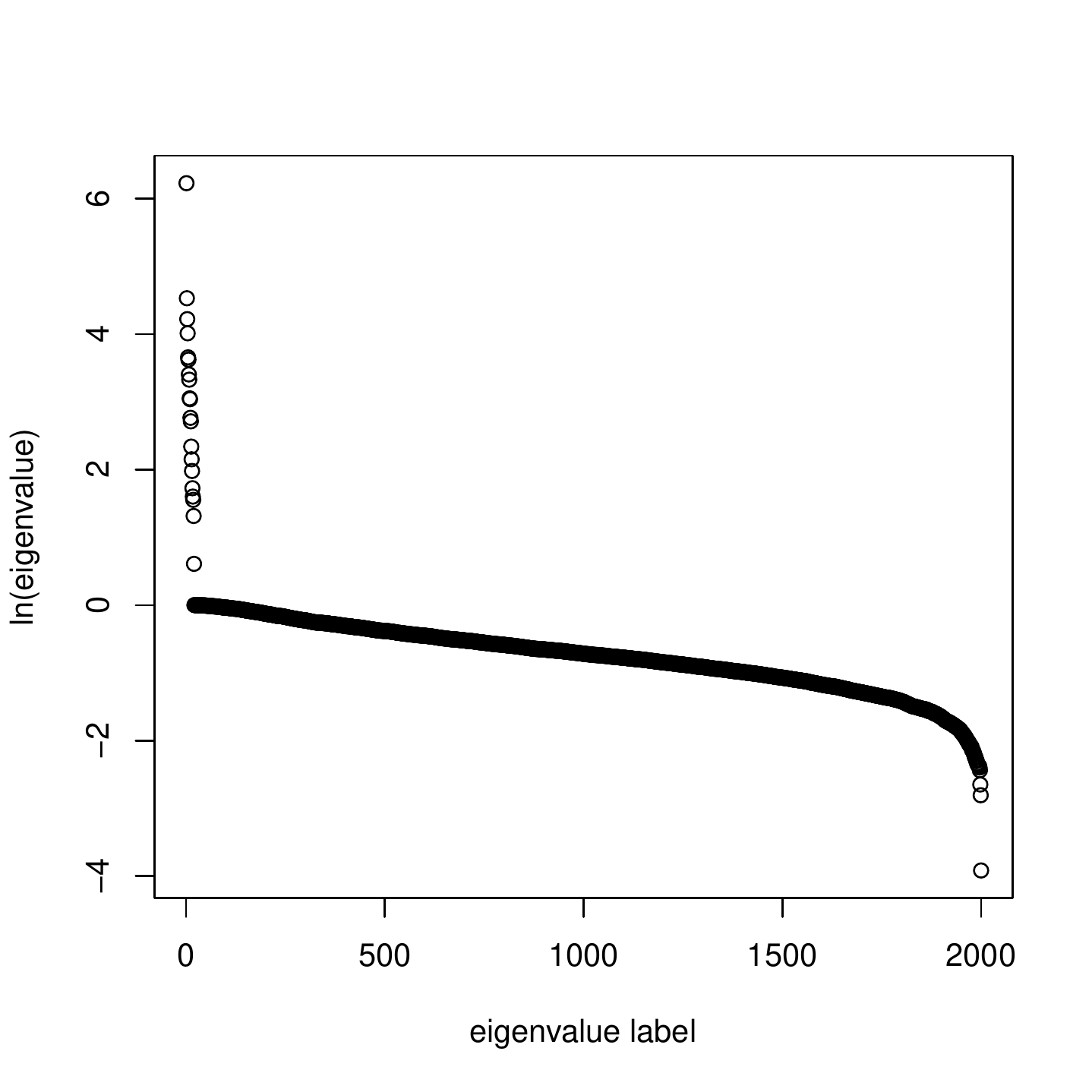}
\caption{A typical graph of the log of the eigenvalues (ordered decreasingly) of the model correlation matrix ${\widetilde\Psi}^{(m)}_{ij}$ for a single ``sampling" ($M=1$). The number of clusters $K=40$. See Subsection \ref{sub.tail} for details.}
\label{Figure3}
\end{figure}

\begin{figure}[ht]
\centering
\includegraphics[scale=1.0]{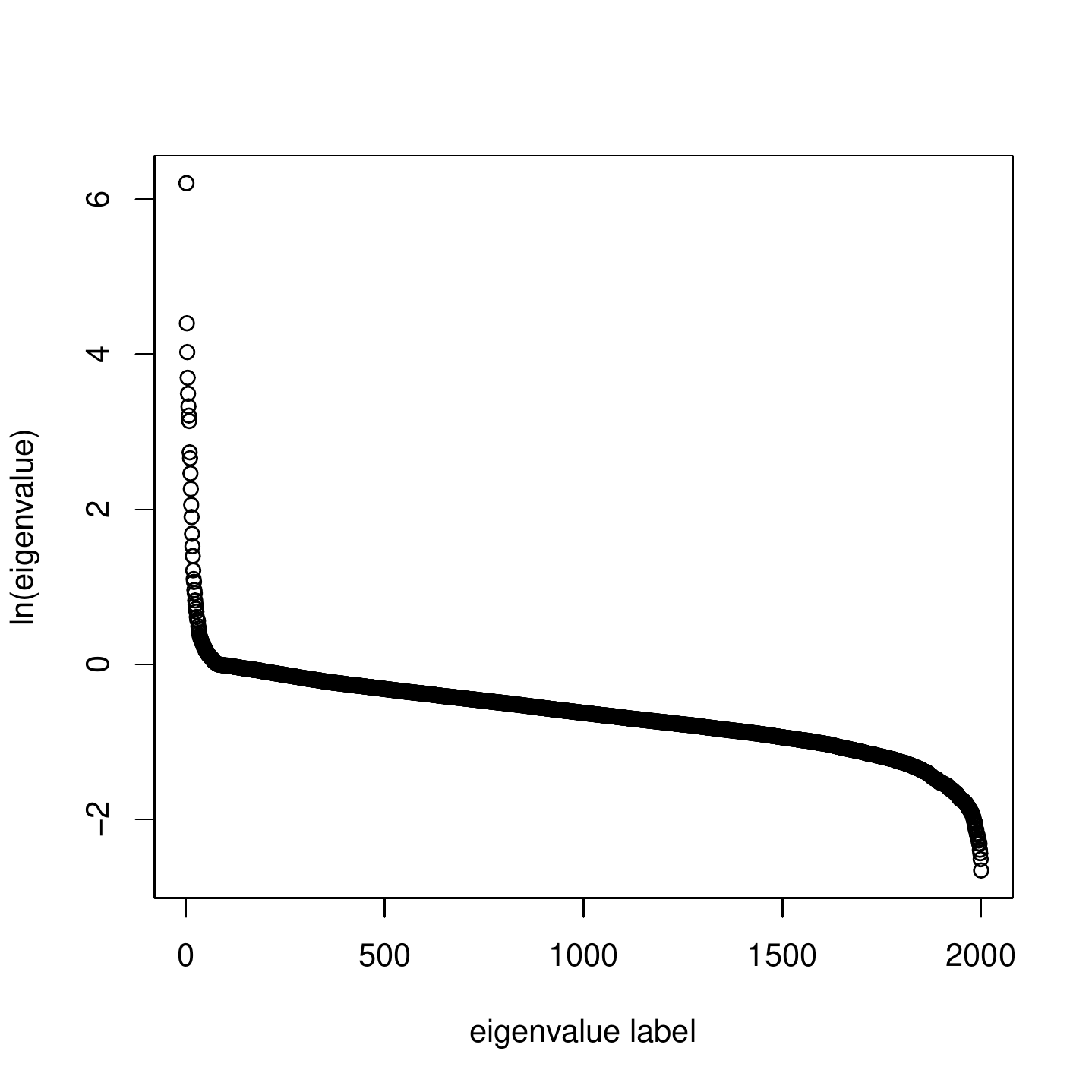}
\caption{A typical graph of the log of the eigenvalues (ordered decreasingly) of the model correlation matrix ${\widetilde\Psi}_{ij}$ obtained by combining $M=100$ ``samplings" (with equal weights). The number of clusters $K=40$. See Subsection \ref{sub.tail} for details.}
\label{Figure4}
\end{figure}

{}Here is a simple method that deals with both of the aforesaid issues at once. To understand this method, it is helpful to look at the eigenvalue graphs given in Figures \ref{Figure1}, \ref{Figure2}, \ref{Figure3}, \ref{Figure4}, which are based on a typical data set of daily returns for $N=2000$ stocks and $T=21$ trading days. These graphs plot the eigenvalues for a single ``sampling" ${\widetilde\Psi}^{(m)}_{ij}$, as well as ${\widetilde\Psi}_{ij}$ based on averaging $M=100$ ``samplings" (with equal weights), for $K=150$ and $K=40$ ($K$ is the number of clusters). Unsurprisingly, there are some small eigenvalues. However, their fraction is small. Furthermore, these small eigenvalues get even smaller for larger values of $K$, but increase when averaging over multiple ``samplings", which also smoothes out the eigenvalue graph structure.

{}What we wish to do is to deform the matrix ${\widetilde\Psi}_{ij}$ by tweaking the small eigenvalues at the tail. We need to define what we mean by the ``tail", i.e., which eigenvalues to include in it. There are many ways of doing this, some are simpler, some are more convoluted. We use a method based on eRank or effective rank \cite{RV}, which can be more generally defined for any subset $S$ of the eigenvalues of a matrix, which (for our purposes here) is assumed to be symmetric and semi-positive-definite. Let
\begin{eqnarray}
 &&\mbox{eRank}(S) = \exp(H)\\
 &&H = -\sum_{a=1}^L p_a~\ln(p_a)\\
 &&p_a = {\lambda^{(a)} \over \sum_{b=1}^L \lambda^{(b)}}
\end{eqnarray}
where $\lambda^{(a)}$ are the $L$ {\em positive} eigenvalues in the subset $S$, and $H$ has the meaning of the (Shannon a.k.a. spectral) entropy \cite{Campbell60}, \cite{YGH}.

{}If we take $S$ to be the full set of $N$ eigenvalues of ${\widetilde \Psi}_{ij}$, then the meaning of $\mbox{eRank}(S)$ is that it is a measure of the effective dimensionality of the matrix ${\widetilde \Psi}_{ij}$. However, this is not what we need to do for our purposes here. This is because the large eigenvalues of ${\widetilde \Psi}_{ij}$ contribute heavily into $\mbox{eRank}(S)$. So, we define $S$ to include all eigenvalues ${\widetilde\lambda}^{(a)}$ ($a=1,\dots,N$) of ${\widetilde \Psi}_{ij}$ that do not exceed 1: $S = \{{\widetilde\lambda}^{(a)} | {\widetilde\lambda}^{(a)} \leq 1\}$. Then we define (here $\mbox{floor}(\cdot) = \lfloor\cdot\rfloor$ can be replaced by $\mbox{round}(\cdot)$)
\begin{equation}\label{eq.eRank}
 n_* = |S| - \mbox{floor}(\mbox{eRank}(S))
\end{equation}
So, the tail is now defined as the set $S_*$ of the $n_*$ smallest eigenvalues ${\widetilde\lambda}^{(a)}$ of ${\widetilde \Psi}_{ij}$.

\begin{figure}[ht]
\centering
\includegraphics[scale=1.0]{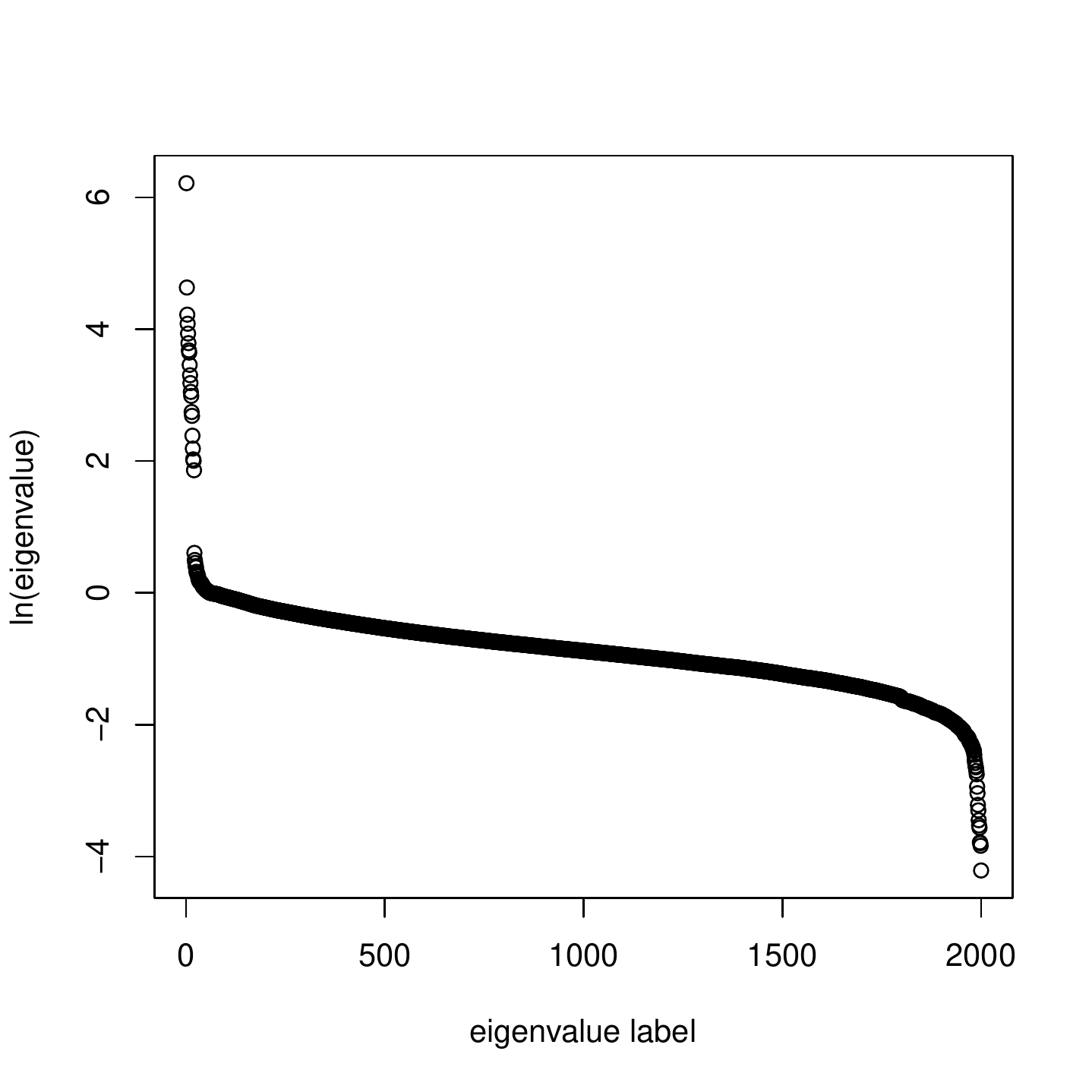}
\caption{A typical graph of the log of the eigenvalues (ordered decreasingly) of the deformed (by adjusting the low-end ``tail" eigenvalues) model correlation matrix ${\widehat \Psi}_{ij}$ obtained by combining $M=100$ ``samplings" (with equal weights). The number of clusters $K=150$. See Subsection \ref{sub.tail} for details.}
\label{Figure5}
\end{figure}

{}We can now deform ${\widetilde\Psi}_{ij}$ by (i) replacing the $n_*$ tail eigenvalues in $S_*$ by ${\widetilde\lambda}_* = \mbox{max}(S_*)$, and (ii) then correcting for the fact that the so-deformed matrix no longer has a unit diagonal. The resulting matrix ${\widehat\Psi}_{ij}$ is given by:
\begin{eqnarray}
 &&{\widehat\Psi}_{ij} = \sum_{a=1}^{N - n_*} {\widetilde \lambda}^{(a)}~{\widetilde V}_i^{(a)}~{\widetilde V}_j^{(a)} + z_i~z_j \sum_{a = N - n_* + 1}^N {\widetilde \lambda}_*~{\widetilde V}_i^{(a)}~{\widetilde V}_j^{(a)}\\
 &&z_i^{2} = y_i^{-2}~\sum_{a = N - n_* + 1}^N {\widetilde \lambda}^{(a)}~[{\widetilde V}_i^{(a)}]^2\\
 &&y_i^2 = \sum_{a = N - n_* + 1}^N {\widetilde \lambda}_*~[{\widetilde V}_i^{(a)}]^2
\end{eqnarray}
Here ${\widetilde V}_i^{(a)}$ are the principal components of ${\widetilde \Psi}_{ij}$. This method is similar to that of \cite{RJ}. The key difference is that in \cite{RJ} the ``adjustments" $z_i$ are applied to all principal components, while here they are only applied to the tail principal components (for which the eigenvalues are deformed). This results in a smaller distortion of the original matrix. The resultant deformed matrix ${\widehat\Psi}_{ij}$ has improved tail behavior (see Figure \ref{Figure5}). Another bonus is that, while superfluously we only modify the tail, the eigenvectors of the deformed matrix ${\widehat\Psi}_{ij}$ are no longer ${\widetilde V}_i^{(a)}$ for all values of $a$, and the eigenvalues outside of the tail are also deformed. In particular, in some cases there can be some (typically, a few) nearly degenerate\footnote{\, They are not degenerate even within the machine precision. However, they are spaced much more closely than other eigenvalues (on average, that is).} eigenvalues ${\widetilde \lambda}^{(a)}$ in the densely populated region of ${\widetilde \lambda}^{(a)}$ (where they are of order 1), i.e., outside of the tail and the higher-end upward-sloping ``neck". The deformation splits such nearly degenerate eigenvalues, which is a welcome bonus. Indeed, the issue with nearly degenerate eigenvalues is that they can adversely affect convergence of the bounded optimization (see below) as the corresponding directions in the risk space have almost identical risk profiles.

\section{Backtests}\label{sec3}

{}Here we discuss some backtests. We wish to see how our machine learning risk models compare with other constructions (see below). For this comparison, we run our backtests exactly as in \cite{Het}, except that the model covariance matrix is build as above (as opposed to the full heterotic risk model construction of \cite{Het}). To facilitate the comparisons, the historical data we use in our backtests here is the same as in \cite{Het}\footnote{\, The same data is also used in \cite{StatIC}, \cite{StatRM}.} and is described in detail in Subsections 6.2 and 6.3 thereof. The trading universe selection is described in Subsection 6.2 of \cite{Het}. We assume that i) the portfolio is established at the open with fills at the open prices; and ii) it is liquidated at the close on the same day (so this is a purely intraday strategy) with fills at the close prices (see \cite{MeanRev} for pertinent details). We include strict trading bounds
\begin{equation}
 |H_i| \leq 0.01~A_i
\end{equation}
Here $H_i$ are the portfolio stock holdings ($i=1,\dots,N$), and $A_i$ are the corresponding historical average daily dollar volumes computed as in Subsection 6.2 of \cite{Het}. We further impose strict dollar-neutrality on the portfolio, so that
\begin{equation}
 \sum_{i=1}^N H_i = 0
\end{equation}
The total investment level in our backtests here is $I$ = \$20M (i.e., \$10M long and \$10M short), same as in \cite{Het}. For the Sharpe ratio optimization with bounds we use the R function {\tt{\small bopt.calc.opt()}} in Appendix C of \cite{Het}. Table \ref{table.summary} gives summaries of the eigenvalues for various values of $K$. Considering that the algorithm is nondeterministic, the results are stable against reruns. Table \ref{table.backtests} summarizes the backtest results. Here we can wonder whether the following would produce an improvement. Suppose we start from the sample correlation matrix $\Psi_{ij}$ and run the algorithm, which produces the model correlation matrix ${\widetilde\Psi}_{ij}$. Suppose now we rerun the algorithm (with the same number of ``samplings" $M$) but use ${\widetilde\Psi}_{ij}$ instead of $\Psi_{ij}$ in Eq. (\ref{Phi}) to build ``sampling" correlation matrices $\Psi^{(m)}_{ij}$. In fact, we can do this iteratively, over and over again, which we refer to as multiple iterations in Table \ref{table.iter}. The results in Table \ref{table.iter} indicate that we do get some improvement on the second iteration, but not beyond. Let us note that for $K\geq 100$ with iterations (see Table \ref{table.iter}) the method of Subsection \ref{sub.tail} was insufficient to deal with the issues with small and nearly degenerate eigenvalues, so we used the full method of \cite{RJ} instead (see Subsection \ref{sub.tail} and Table \ref{table.iter} for details), which distorts the model correlation matrix more (and this affects performance).

\section{Concluding Remarks}\label{sec4}

{}So, the machine learning risk models we discuss in this paper outperform statistical risk models \cite{StatRM}. They have the performance essentially similar to the heterotic risk models based on statistical industry classifications using multilevel clustering \cite{StatIC}. However, here we have single-level clustering, and there is no aggregation of clusterings as in \cite{StatIC}. Also, the resultant model correlation matrix ${\widetilde \Psi}_{ij}$ is {\em not} a factor model, whereas the models of \cite{StatIC} are factor models. Note that both the machine learning risk models of this paper and the models of \cite{StatIC} still underperform the heterotic risk models based on fundamental industry classifications; see \cite{Het}, \cite{HetPlus}.

{}In this regard, let us tie up a few ``loose ends", so to speak. Suppose we take just a single ``sampling" $\Psi^{(m)}_{ij}$. This is an incomplete, single-level heterotic risk model. However, $\Psi^{(m)}_{ij}$ by construction is positive-definite, so we can invert it and use it in optimization. So, does averaging over a large number $M$ of ``samplings" (as in the machine learning risk models of this paper), or implementing a multilevel ``Russian-doll" embedding \cite{RD} as in \cite{StatIC}, add value? It does. Thus, two runs based on a single ``sampling" with $K=40$ and $M=1$ produced the following results: (i) ROC = 42.434\%, SR = 15.479, CPS = 2.044; and (ii) ROC = 42.735\%, SR = 15.51, CPS = 2.054 (see Table \ref{table.backtests} for notations). Also, what if, instead of using a single k-means to compute $\Psi^{(m)}_{ij}$, we aggregate a large number $P$ of k-means clusterings as in \cite{StatIC}? This does not appear to add value. Here are the results from a typical run with $K=30$, $M=100$ and $P=100$: ROC = 42.534\%, SR = 15.764, CPS = 2.09. Apparently, and perhaps unsurprisingly, aggregating multiple clusterings and averaging over multiple ``samplings" has similar effects. This, in fact, is reassuring.

\appendix
\section{R Code}\label{app.A}

{}In this appendix we give R (R Project for Statistical Computing, \url{https://www.r-project.org/}) source code for constructing machine learning risk models discussed in the main text. The code is straightforward and self-explanatory. The sole function is {\tt{\small qrm.calc.ml.cor.mat()}} with the following inputs: {\tt{\small r1}} is the $N \times T$ matrix of returns ($N$ is the number of stocks, $T$ is the number of points in the time series); {\tt{\small k}} is the number of clusters $K$; {\tt{\small nn}} is the number of iterations (see Section \ref{sec3}); {\tt{\small calc.num}} is the number of ``samplings" $M$; {\tt{\small iter.num}} is the maximum number of iterations (which we always set to 100, and which was never saturated in any of our hundreds of thousands of {\tt{\small kmeans()}} calls) used by the built-in R function {\tt{\small kmeans()}} internally called via the function {\tt{\small qrm.stat.ind.class()}} given in Appendix A of \cite{StatIC}; {\tt{\small num.try}} is the number of clusterings {\tt{\small qrm.stat.ind.class()}} aggregates internally, with {\tt{\small num.try = 1}} (which is the value we use) corresponding to a single k-means clustering; {\tt{\small reg.tail}} is the Boolean for regularizing (when set to {\tt{\small TRUE}}) the tail of the eigenvalues as in Subsection \ref{sub.tail}. The output of {\tt{\small qrm.calc.ml.cor.mat()}} is the inverse $\Gamma_{ij}^{-1}$ of the model covariance matrix $\Gamma_{ij} = \sigma_i\sigma_j{\widetilde\Psi}_{ij}$ (or $\Gamma_{ij} = \sigma_i\sigma_j{\widehat\Psi}_{ij}$, when {\tt{\small reg.tail = TRUE}} -- see Subsection \ref{sub.tail}), where $\sigma_i^2$ are the sample variances. The weights with which the $M$ ``samplings" are combined are internally set to be uniform. However, this can be modified if so desired. The weights can be based on the Euclidean or some other distance, the sum over the specific variances $\xi_i^2$, the average correlations, etc. In our simulations nontrivial weights did not add value.\\
\\
{\tt{\small
\noindent qrm.calc.ml.cor.mat <- function (r1, k, nn = 1, \\
\indent calc.num = 100, iter.max = 100, \\
\indent num.try = 1, reg.tail = F) \\
\{\\
\indent calc.mod.erank <- function(x)\\
\indent \{\\
\indent \indent take <- log(x) > 0\\
\indent \indent n <- sum(take)\\
\indent \indent x <- x[!take]\\
\indent \indent p <- x / sum(x)\\
\indent \indent h <- - sum(p * log(p))\\
\indent \indent er <- exp(h)\\
\indent \indent er <- er + n\\
\indent \indent return(er)\\
\indent \}\\
\\
\indent calc.het.cor <- function(p, ind)\\
\indent \{\\
\indent \indent u <- rep(0, nrow(ind))\\
\indent \indent for(a in 1:ncol(ind))\\
\indent \indent \{\\
\indent \indent \indent tt <- ind[, a] == 1\\
\indent \indent \indent p1 <- p[tt, tt]\\
\indent \indent \indent p1 <- eigen(p1)\\
\indent \indent \indent u[tt] <- p1\$vectors[, 1]\\
\indent \indent \}\\
\indent \indent flm <- u * ind\\
\indent \indent q <- t(flm) \%*\% p \%*\% flm\\
\indent \indent g <- flm \%*\% q \%*\% t(flm)\\
\indent \indent diag(g) <- 1\\
\indent \indent return(g)\\
\indent \}\\
\\
\indent calc.cor.mat <- function(p, r1, k, iter.max, num.try)\\
\indent \{\\
\indent \indent ww <- 0\\
\indent \indent gg <- 0\\
\indent \indent for(j in 1:calc.num)\\
\indent \indent \{\\
\indent \indent \indent ind <- qrm.stat.ind.class(r1, k, \\
\indent \indent \indent \indent iter.max = iter.max, num.try = num.try, \\
\indent \indent \indent \indent demean.ret = F)\\
\indent \indent \indent g <- calc.het.cor(p, ind)\\
\indent \indent \indent w <- 1 \#\#\# uniform weighting\\
\indent \indent \indent gg <- gg + g * w\\
\indent \indent \indent ww <- ww + w\\
\indent \indent \}\\
\indent \indent gg <- gg / ww\\
\indent \indent return(gg)\\
\indent \}\\
\\
\indent gg <- cor(t(r1), t(r1))\\
\indent for(a in 1:nn)\\
\indent \indent gg <- calc.cor.mat(gg, r1, k, iter.max, num.try)\\
\indent if(reg.tail)\\
\indent \{\\
\indent \indent xx <- eigen(gg)\\
\indent \indent vv <- xx\$values\\
\indent \indent uu <- xx\$vectors\\
\indent \indent er <- trunc(calc.mod.erank(vv))\\
\indent \indent tt <- (er + 1):length(vv)\\
\indent \indent zz <- colSums(t(uu[, tt]\^{}2) *\\
\indent \indent \indent vv[tt]) / vv[er] / rowSums(uu[, tt]\^{}2)\\
\indent \indent zz <- sqrt(zz)\\
\indent \indent vv[tt] <- vv[er]\\
\indent \indent uu <- t(t(uu) * sqrt(vv))\\
\indent \indent uu[, tt] <- zz * uu[, tt]\\
\indent \indent gg <- uu \%*\% t(uu)\\
\indent \}\\
\indent gg <- solve(gg)\\
\indent ss <- apply(r1, 1, sd)\\
\indent gg <- t(gg / ss) / ss\\
\indent return(gg)\\
\}
}}

\section{DISCLAIMERS}\label{app.B}

{}Wherever the context so requires, the masculine gender includes the feminine and/or neuter, and the singular form includes the plural and {\em vice versa}. The author of this paper (``Author") and his affiliates including without limitation Quantigic$^\circledR$ Solutions LLC (``Author's Affiliates" or ``his Affiliates") make no implied or express warranties or any other representations whatsoever, including without limitation implied warranties of merchantability and fitness for a particular purpose, in connection with or with regard to the content of this paper including without limitation any code or algorithms contained herein (``Content").

{}The reader may use the Content solely at his/her/its own risk and the reader shall have no claims whatsoever against the Author or his Affiliates and the Author and his Affiliates shall have no liability whatsoever to the reader or any third party whatsoever for any loss, expense, opportunity cost, damages or any other adverse effects whatsoever relating to or arising from the use of the Content by the reader including without any limitation whatsoever: any direct, indirect, incidental, special, consequential or any other damages incurred by the reader, however caused and under any theory of liability; any loss of profit (whether incurred directly or indirectly), any loss of goodwill or reputation, any loss of data suffered, cost of procurement of substitute goods or services, or any other tangible or intangible loss; any reliance placed by the reader on the completeness, accuracy or existence of the Content or any other effect of using the Content; and any and all other adversities or negative effects the reader might encounter in using the Content irrespective of whether the Author or his Affiliates is or are or should have been aware of such adversities or negative effects.

{}The R code included in Appendix \ref{app.A} hereof is part of the copyrighted R code of Quantigic$^\circledR$ Solutions LLC and is provided herein with the express permission of Quantigic$^\circledR$ Solutions LLC. The copyright owner retains all rights, title and interest in and to its copyrighted source code included in Appendix \ref{app.A} hereof and any and all copyrights therefor.

\newpage

\begin{table}[ht]
\caption{Summary of eigenvalues of the model correlation matrix ${\widetilde \Psi}_{ij}$ for the indicated values of the number $K$ of clusters. All runs are for the number of ``samplings" $M=100$ except for the second entry with $K=100$ marked with an asterisk, for which $M=1000$. 1st Qu. = first quartile; 3rd Qu. = third quartile. Mean is always 1 as ${\widetilde \Psi}_{ij}$ is a correlation matrix with a unit diagonal (and the sum of eigenvalues equals the sum of the diagonal elements).} 
\begin{tabular}{l l l l l l l} 
\hline\hline 
$K$ & Min & 1st Qu. & Median & Mean & 3rd Qu. & Max\\[0.5ex] 
\hline 
10 & 0.078 & 0.4795 & 0.6579 & 1 & 0.8318 & 514.9 \\
20 & 0.0684 & 0.45 & 0.6114 & 1 & 0.7856 & 503.6 \\
30 & 0.0695 & 0.4221 & 0.5662 & 1 & 0.7533 & 499.7 \\
40 & 0.07 & 0.3895 & 0.5346 & 1 & 0.7295 & 497.8 \\
50 & 0.0684 & 0.3722 & 0.515 & 1 & 0.7025 & 496 \\
60 & 0.0685 & 0.3574 & 0.4979 & 1 & 0.6841 & 495.9 \\
70 & 0.0661 & 0.3469 & 0.4838 & 1 & 0.6686 & 497.1 \\
70 & 0.0665 & 0.3477 & 0.4848 & 1 & 0.6701 & 496.9 \\
70 & 0.0653 & 0.3464 & 0.483 & 1 & 0.6686 & 496.9 \\
70 & 0.0652 & 0.3467 & 0.4825 & 1 & 0.6663 & 497.4 \\
70 & 0.0642 & 0.3474 & 0.4835 & 1 & 0.67 & 496.6 \\
70 & 0.0662 & 0.3477 & 0.4843 & 1 & 0.6679 & 496.7 \\
70 & 0.064 & 0.3473 & 0.4853 & 1 & 0.6691 & 496.9 \\
80 & 0.0614 & 0.3393 & 0.4739 & 1 & 0.6532 & 497.6 \\
90 & 0.0355 & 0.3298 & 0.4626 & 1 & 0.641 & 497.6 \\
100 & 0.015 & 0.3241 & 0.4532 & 1 & 0.6307 & 498.3 \\
100$^*$ & 0.0152 & 0.3318 & 0.4618 & 1 & 0.6276 & 498.1 \\
101 & 0.0184 & 0.3217 & 0.4515 & 1 & 0.6278 & 498.6 \\
102 & 0.0203 & 0.3219 & 0.4512 & 1 & 0.6255 & 498.6 \\
103 & 0.0197 & 0.321 & 0.4496 & 1 & 0.6268 & 498.4 \\
104 & 0.0153 & 0.319 & 0.4482 & 1 & 0.6245 & 498.5 \\
105 & 0.0088 & 0.3204 & 0.4491 & 1 & 0.6236 & 498.2 \\
106 & 0.0116 & 0.3191 & 0.447 & 1 & 0.6213 & 498.4 \\
107 & 0.009 & 0.3176 & 0.4466 & 1 & 0.6215 & 498.2 \\
108 & 0.0067 & 0.3165 & 0.4441 & 1 & 0.6187 & 498.4 \\
109 & 0.0105 & 0.319 & 0.4447 & 1 & 0.6182 & 498.2 \\
110 & 0.0032 & 0.3149 & 0.4432 & 1 & 0.6176 & 498.3 \\
120 & 0.0026 & 0.3103 & 0.4355 & 1 & 0.6081 & 499 \\
130 & 0.0051 & 0.3023 & 0.4259 & 1 & 0.5986 & 499.7 \\
140 & 0.0022 & 0.2976 & 0.4209 & 1 & 0.5917 & 499.4 \\
150 & 0.002 & 0.292 & 0.4132 & 1 & 0.5839 & 499.8 \\[1ex] 
\hline 
\end{tabular}
\label{table.summary} 
\end{table}

\begin{table}[ht]
\caption{Backtest results for machine learning risk models for the indicated number $K$ of clusters (here and in Tables \ref{table.iter} and \ref{table.RJ} the number of ``samplings" $M = 100$). ROC = annualized Return-on-Capital (in \%). SR = annualized daily Sharpe Ratio \cite{Sharpe1994}. CPS = Cents-per-Share. The cases marked ``tail" correspond to using the deformed model correlation matrix ${\widehat\Psi}_{ij}$; see Subsection \ref{sub.tail} for details. Also see Figures \ref{Figure6}, \ref{Figure7}, \ref{Figure8} for graphs of ROC, SR and CPS based on these results.} 
\begin{tabular}{l l l l} 
\hline\hline 
$K$ & ROC (\%) & SR & CPS\\[0.5ex] 
\hline 
10 & 42.643 & 15.524 & 2.059 \\
20 & 43.135 & 16.089 & 2.093 \\
30 & 43.11 & 16.337 & 2.095 \\
40 & 43.025 & 16.409 & 2.094 \\
40 & 42.961 & 16.366 & 2.091 \\
50 & 42.895 & 16.43 & 2.091 \\
50 & 42.891 & 16.486 & 2.091 \\
60 & 42.647 & 16.414 & 2.084 \\
70 & 42.449 & 16.358 & 2.08 \\
80 & 42.131 & 16.313 & 2.071 \\
90 & 41.842 & 16.236 & 2.064 \\
100 & 41.387 & 16.096 & 2.051 \\
110 & 40.958 & 16.057 & 2.041 \\
120, tail  & 40.726 & 15.902 & 2.033 \\
130, tail  & 40.215 & 15.838 & 2.019 \\
140, tail  & 39.894 & 15.819 & 2.011 \\
150, tail  & 39.162 & 15.668 & 1.986 \\[1ex] 
\hline 
\end{tabular}
\label{table.backtests} 
\end{table}

\begin{table}[ht]
\caption{Backtest results for machine learning risk models for the indicated number $K$ of clusters with iterations (see Section \ref{sec3} for details). X2, X3, X4 stand for 2, 3, 4 iterations, respectively. The cases marked ``tail" correspond to using the deformed model correlation matrix ${\widehat\Psi}_{ij}$ based on the method of \cite{RJ} (and {\em not} on the method of Subsection \ref{sub.tail}), whereby ${\widehat\Psi}_{ij} = \Theta_{ij} / \sqrt{\Theta_{ii}} \sqrt{\Theta_{jj}}$, $\Theta_{ij} = \sum_{a=1}^{N - n_*} {\widetilde \lambda}^{(a)}~{\widetilde V}_i^{(a)}~{\widetilde V}_j^{(a)} + \sum_{a = N - n_* + 1}^N {\widetilde \lambda}_*~{\widetilde V}_i^{(a)}~{\widetilde V}_j^{(a)}$; see Subsection \ref{sub.tail} for notations. For comparison purposes, also see Table \ref{table.RJ}, which gives backtest results for machine learning risk models {\em without} iterations using the deformed model correlation matrix ${\widehat\Psi}_{ij}$ based on the method of \cite{RJ}.} 
\begin{tabular}{l l l l} 
\hline\hline 
$K$ & ROC (\%) & SR & CPS\\[0.5ex] 
\hline 
10, X2 & 42.614 & 15.213 & 2.036 \\
10, X2 & 42.609 & 15.204 & 2.036 \\
10, X2 & 42.627 & 15.236 & 2.037 \\
20, X2 & 43.468 & 15.82 & 2.087 \\
30, X2 & 43.64 & 16.054 & 2.099 \\
40, X2 & 43.672 & 16.207 & 2.102 \\
40, X2 & 43.668 & 16.186 & 2.102 \\
40, X3 & 43.643 & 16.026 & 2.091 \\
40, X4 & 43.508 & 15.899 & 2.08 \\
50, X2 & 43.676 & 16.296 & 2.103 \\
60, X2 & 43.75 & 16.398 & 2.109 \\
70, X2 & 43.714 & 16.396 & 2.112 \\
80, X2 & 43.62 & 16.41 & 2.113 \\
90, X2 & 43.501 & 16.418 & 2.113 \\
90, X3 & 43.654 & 16.33 & 2.109 \\
90, X4 & 43.537 & 16.22 & 2.097 \\
100, X2, tail & 43.216 & 16.213 & 2.086 \\
110, X2, tail & 43.153 & 16.198 & 2.087 \\
120, X2, tail & 43.091 & 16.244 & 2.088 \\
130, X2, tail & 43.001 & 16.214 & 2.089 \\
140, X2, tail & 42.944 & 16.249 & 2.09 \\
150, X2, tail & 42.91 & 16.267 & 2.093 \\[1ex] 
\hline 
\end{tabular}
\label{table.iter} 
\end{table}

\begin{table}[ht]
\caption{Backtest results for machine learning risk models for the indicated number $K$ of clusters, all {\em without} iterations, and using the deformed model correlation matrix ${\widehat\Psi}_{ij}$ based on the method of \cite{RJ}; see Table \ref{table.iter} for details.} 
\begin{tabular}{l l l l} 
\hline\hline 
$K$ & ROC (\%) & SR & CPS\\[0.5ex] 
\hline 
10 & 42.421 & 15.337 & 2.033 \\
20 & 42.824 & 15.853 & 2.062 \\
30 & 42.845 & 16.027 & 2.066 \\
40 & 42.805 & 16.187 & 2.068 \\
50 & 42.567 & 16.186 & 2.06 \\
60 & 42.425 & 16.171 & 2.057 \\
70 & 42.263 & 16.121 & 2.054 \\
80 & 42.191 & 16.155 & 2.055 \\
90 & 42.096 & 16.117 & 2.054 \\
100 & 41.854 & 16.079 & 2.047 \\
110 & 41.645 & 15.991 & 2.041 \\
120 & 41.431 & 15.967 & 2.035 \\
130 & 41.44 & 15.966 & 2.04 \\
140 & 41.344 & 15.989 & 2.038 \\
150 & 41.156 & 15.995 & 2.033 \\[1ex] 
\hline 
\end{tabular}
\label{table.RJ} 
\end{table}

\newpage\clearpage

\newpage\clearpage

\newpage\clearpage

\newpage\clearpage

\newpage\clearpage

\newpage\clearpage
\begin{figure}[ht]
\centering
\includegraphics[scale=1.0]{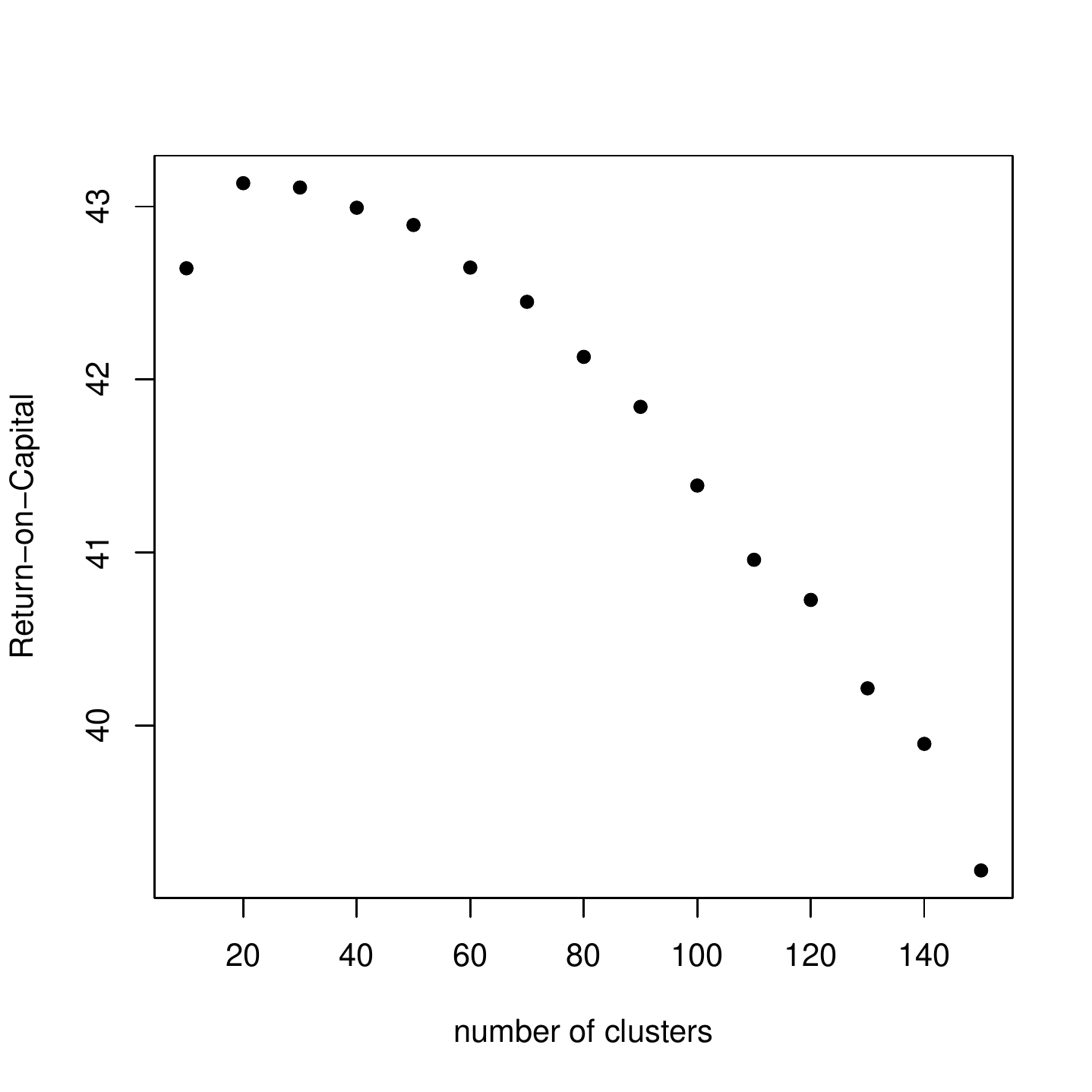}
\caption{Return-on-Capital (ROC) vs. $K$ (the number of clusters) based on simulations from Table \ref{table.backtests}. For multiple simulations per $K$, the average ROC is shown.}
\label{Figure6}
\end{figure}

\newpage\clearpage
\begin{figure}[ht]
\centering
\includegraphics[scale=1.0]{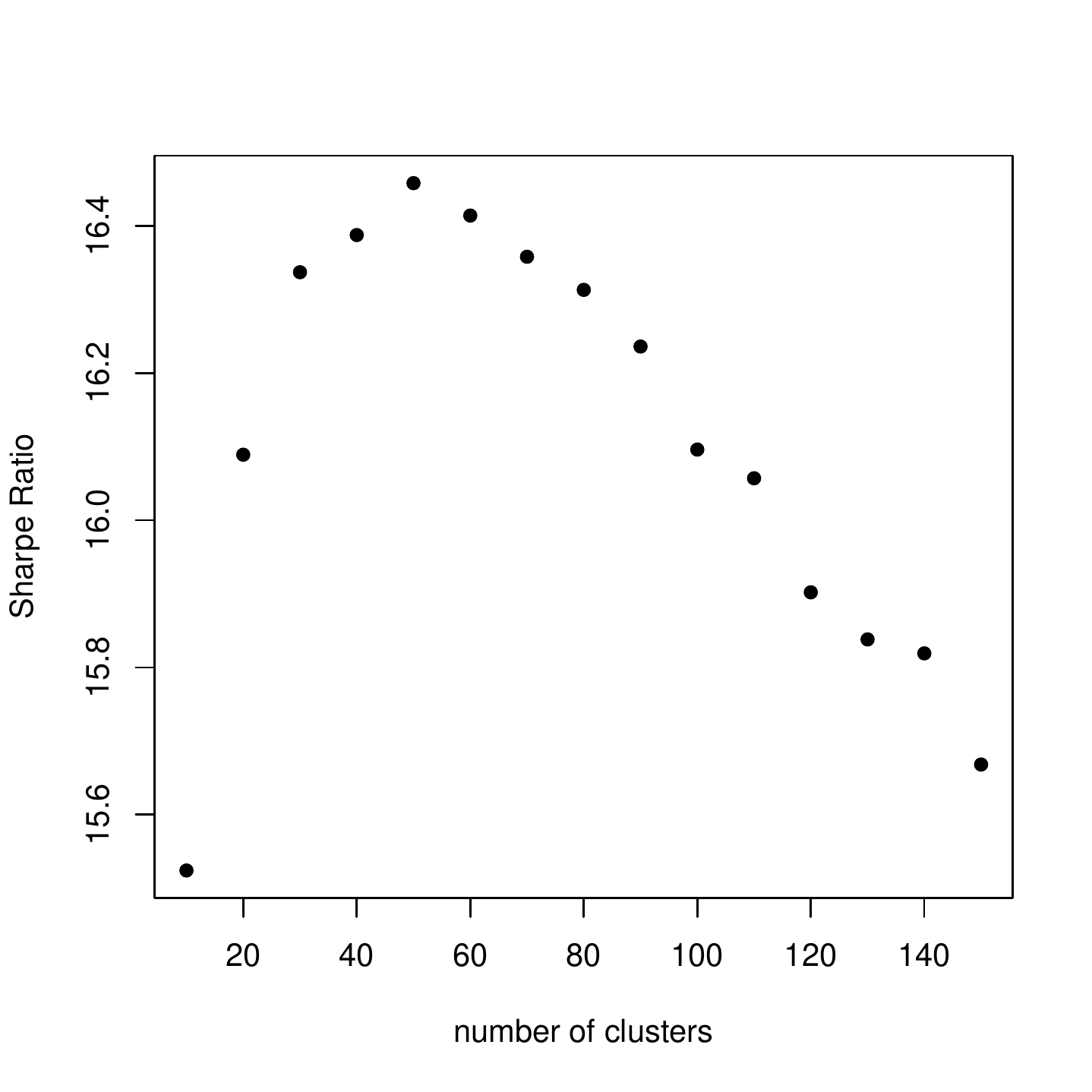}
\caption{Sharpe Ratio (SR) vs. $K$ (the number of clusters) based on simulations from Table \ref{table.backtests}. For multiple simulations per $K$, the average SR is shown.}
\label{Figure7}
\end{figure}

\newpage\clearpage
\begin{figure}[ht]
\centering
\includegraphics[scale=1.0]{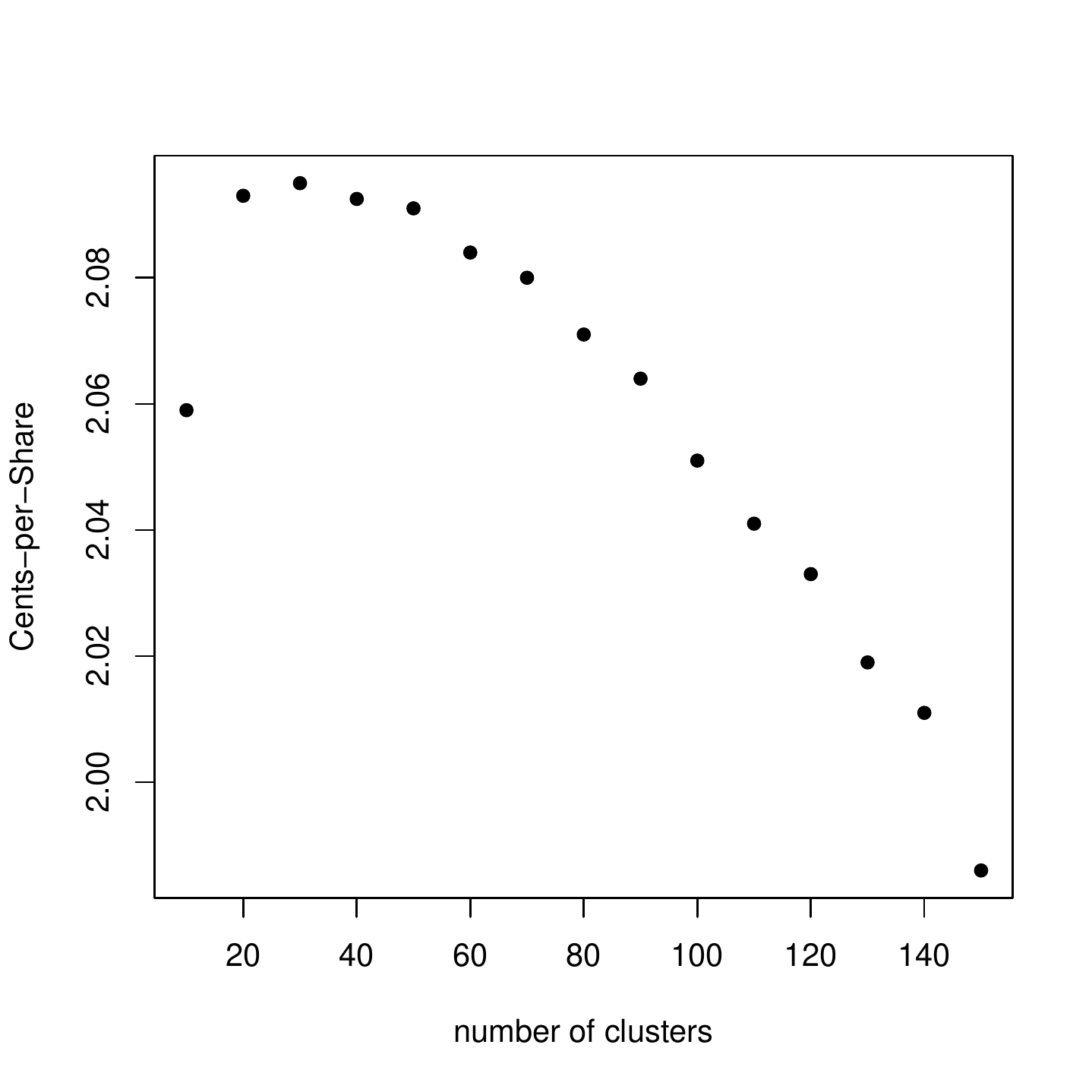}
\caption{Cents-per-Share (CPS) vs. $K$ (the number of clusters) based on simulations from Table \ref{table.backtests}. For multiple simulations per $K$, the average CPS is shown.}
\label{Figure8}
\end{figure}

\end{document}